\documentclass[aps,preprint,prb,showpacs,amsmath,amssymb]{revtex4}
\usepackage{graphicx}
\usepackage{dcolumn}
\usepackage{bm}
\usepackage{float}
\begin{document}
\title{Density-functional theory of nonequilibrium tunneling}
\author{Per Hyldgaard}
\affiliation{Department of Microtechnology and Nanoscience, 
MC2, Chalmers University of Technology, 
SE--41296 Gothenburg, Sweden}
\date{August 3, 2008}

\begin{abstract}
Nanoscale optoelectronics and molecular-electronics systems 
operate with current injection and nonequilibrium tunneling,
phenomena that challenge consistent descriptions of the 
steady-state transport.  
The current affects the electron-density variation and hence the 
inter- and intra-molecular bonding which in turn determines the 
transport magnitude.  The standard approach for efficient 
characterization of steady-state tunneling 
combines ground-state density functional theory (DFT) calculations 
(of an effective scattering potential) with a Landauer-type 
formalism and ignores all actual many-body scattering.
The standard method also lacks a formal variational basis.
This paper formulates a Lippmann-Schwinger collision density 
functional theory (LSC-DFT) for tunneling transport with full 
electron-electron interactions. Quantum-kinetic (Dyson) equations 
are used for an exact reformulation that expresses the variational 
noninteracting and interacting many-body scattering T-matrices in 
terms of universal density functionals. The many-body
Lippmann-Schwinger (LS) variational principle defines
an implicit equation for the exact nonequilibrium density. 

\end{abstract}

\maketitle

\section{Introduction} 
The function of heterostructure-based optoelectronics 
and of  future molecular electronics~\cite{MolEl} rests on 
current-injection and tunneling transport that causes 
genuinely nonequilibrium conditions. The systems are 
nanoscale and the performance is sensitive to the atomic 
configuration.  The technologies define difficult theory 
problems of calculating
nonequilibrium tunneling in the presence of electron-electron/boson
scattering. A quantum cascade laser~\cite{QCL} (QCL) produces 
optical transitions in repeated stages with current injection in 
resonant-tunneling structures.  The electron-electron interaction 
is ubiquitous, causes strong many-body scattering out of 
equilibrium,\cite{QCLdisc} and directly affects the QCL operation. 
This is because electrons which exit one QCL stage must be recycled 
for current injection in subsequent stages and because the optical 
activity depends on details of the energy distribution of injected 
electrons.  Similarly, a current-driven molecular-electronics 
switch~\cite{MolSwitch} positions a nanoscale molecule between 
leads and uses charge-transfer processes to adjust the intra- 
and inter-molecular bonding and morphology.  Understanding details of 
the current-induced changes in the interacting electron distribution 
is also here important because the molecular morphology (and nature of 
bonds) determines the magnitude of the nonequilibrium tunneling.  
Deriving a quantum-physical description that provides {\it ab 
initio,} predictive and self-consistent accounts of nonequilibrium 
tunneling with full electron-electron interaction is very desirable. 

For equilibrium systems it is possible to rely on the traditional,
{\it ground-state\/} density functional theory~\cite{HohKohn,KohnSham} 
(DFT) to provide a materials-/system-specific account. The ground-state 
DFT is formulated in the canonical ensemble, conserves the total number 
of electrons, and takes as input only the potential defined by the nuclei.
The approach uses predefined approximations for a universal 
functional that expresses contributions of the electron-electron 
interaction to the total, interacting ground-state energy for a
system in equilibrium.  Traditional implementations of 
ground-state DFT calculations,
using the local density approximation~\cite{LDA} and/or the
generalized-gradient approximation,\cite{GGA} provide a rich 
description of the bonding inside materials with a dense 
electron-density distribution and within molecules.  Recent 
extensions~\cite{vdWDFother,vdWDF,vdWDFsc} of the 
density functionals to include accounts of dispersive interactions 
allow descriptions of vdW-DF bonding and organization in sparse 
materials~\cite{graphitics} and in (as well as of) 
macromolecules.\cite{molsys} The ground-state DFT calculations 
define an {\it ab initio level\/} (as opposed to a model level)
of details in the description of materials-/molecular-physics 
properties and are extremely useful for they allow a transferable 
description of specific systems. 
It is this {\it ab initio level\/} of detail that we seek also for 
the open, nonequilibrium tunneling systems. Unfortunately, the 
openness and transport invalidate the assumptions of 
traditional DFT which rests in a {\it ground-state-total-energy\/} 
variation principle that applies in equilibrium.

There is exciting recent progress towards {\it ab initio\/} calculations 
of steady-state interacting tunneling even if consistent quantum-physical 
calculations of open nonequilibrium systems are challenging.  The 
nonequilibrium tunneling problem is difficult because the finite 
applied bias makes phase space available for actual electron scattering 
processes (which are normally suppressed in equilibrium).  It is also 
difficult because it is essential (but hard) to ensure conservation 
rules~\cite{KadBaym} in accounts of the nonequilibrium and interacting 
electron dynamics.  
The Landauer-type formulations,\cite{LandauerButtiker} 
characterizing tunneling transport in terms of ballistic propagation 
of individual particles moving in an effective potential, constitute
a simple approximation. They can formally be 
derived~\cite{Baranger,KondoPRL,NedYigal} for the linear-response 
regime or in the absence of many-body scattering using nonequilibrium 
Green functions~\cite{NEQGF,Langreth} (NEGF) within the so-called partition 
scheme introduced by Caroli {\it et al},\cite{Caroli} but they only have 
limited applicability.  One solution strategy for {\it ab 
initio\/} calculations of nonequilibrium interacting tunneling involves 
use of quantum-kinetic equations~\cite{Langreth} for conserving 
calculations and there exists a number of formal 
results,\cite{Datta,KondoPRL,NedYigal,RateEq,NEQreform,Stefanucci} 
extending and correcting the Landauer-type formulation.
Several explicit calculations for nonequilibrium tunneling with 
electron-electron interactions focus on correlated-electron 
systems like the Kondo problem and use both diagrammatic approaches 
(for example, Refs.~\onlinecite{KondoPRL,NedLee,NedLeeYigal}) and 
exact reformulations\cite{NEQreform,exactKondo} of the quantum-kinetic 
equations.  Tunneling through molecular systems has been investigated on 
an {\it ab initio level\/} diagrammatically starting from either Hartree-Fock 
eigenstates of the molecular system\cite{Wenzel} or by applying a 
conserving GW formulation.\cite{Thygesen} A second solution strategy 
for interacting steady-state tunneling invokes time-dependent 
density functional theory~\cite{RungeGross} (TD-DFT) either in 
combination with a master-equation approach for a finite, closed-loop 
system~\cite{Burke,GebCar} or with the NEGF formulation in an 
infinite, open, partition-free formulation with an explicit time-evolution 
in the applied bias.\cite{Stefanucci,TDDFTpartitionCalcs} 
Traditional TD-DFT contains no implicit dephasing 
and describes coherent evolution of the (interacting) many-body 
wavefunction for finite systems. Explicit and implicit
dephasing mechanisms must be carefully analyzed~\cite{GebCar,Stefanucci}
in these applications to steady-state tunneling.  Both of the 
TD-DFT-based methods allow calculations of interacting tunneling 
transport that are in principle exact, and the TD-DFT basis should 
make it simpler to achieve an {\it ab initio level\/} of details.
All of the above-mentioned solution strategies for {\it ab initio\/} 
nonequilibrium calculations are computationally intensive.

The standard method for efficient calculations of steady-state 
tunneling with an {\it ab initio level\/} of detail combines 
ground-state DFT with a Landauer-type formulation,\cite{LandauerButtiker}
computing tunneling transport as an independent-particle transmission
problem.  The resulting ballistic-transport DFT (BT-DFT) approach 
was introduced by Lang~\cite{Lang} and uses the 
ground-state DFT exchange-correlation energy to define an 
effective potential for scattering of independent particles.  
The BT-DFT represents a natural (but {\it ad hoc}) extention of 
the Poisson-Landauer-type transport solvers.\cite{Poisson}  It is a
meaningful approximation because electron conservation is automatic 
in the ballistic single-particle description. There exist 
efficient implementations~\cite{DiVentra,TranSiesta} and BT-DFT
calculations provide valuable theory characterizations of tunneling
systems.  However, the BT-DFT approach also constitutes an 
uncontrolled approximation.  Use of a ground-state DFT formulation 
must be discussed even in linear response.\cite{BurkeDisc} Use of the 
Landauer-type formulation, ignoring all actual many-body scattering 
events, is problematic for fully nonequilibrium conditions. 
Analysis~\cite{Phon} of the frequency-resolved current density 
shows that nonequilibrium electron-phonon scattering causes 
distribution changes that are inconsistent with Landauer-type 
descriptions (corresponding problems for nonequilibrium
electron-electron scattering can be inferred from Ref.~\onlinecite{QCLdisc}). 
The use of a Landauer-type formulation prevents BT-DFT from consistent
calculations of current-induced changes in the electron distribution
and hence of effects that are important for nanoscale optoelectronics 
and molecular electronics. The fact that BT-DFT lacks a formal variational basis
motivates a continued search for efficient {\it ab initio\/} calculations 
of nonequilibrium tunneling with full electron-electron interaction.

This paper formulates a Lippmann-Schwinger collision density 
functional theory (LSC-DFT) for steady-state, nonequilibrium 
tunneling systems treated in the Born-Oppenheimer approximation.
The LSC-DFT is based on formal collision theory~\cite{Albert} for 
the interacting many-body problem~\cite{AlbertXIX29,Yang} and 
allows an exact discussion, resting on the Lippmann-Schwinger 
(LS) variational principle.\cite{LippSchwing,Pirenne,Gellmann}
The LSC-DFT is expressed through universal density functionals
that characterize the variational form of the noninteracting 
and interacting many-body T-matrices.  The LSC-DFT provides a 
formal solution in terms of an implicit equation for the exact 
electron density. The formal LSC-DFT solution constitutes
a natural starting point in a search for rigorous formulations 
of single-particle schemes based on the LS variational principle.

This paper is organized as follows. Section II defines the
partition scheme and the Hamiltonian. It also discusses the general
(nonequilibrium and interacting) tunneling problem as a 
complex many-body collision problem.  Section III presents the 
formal density-functional basis for the theory while section IV 
develops the LS T-matrix functional description.
Finally, section V contains a summary and outlook while 
appendix A provides details of the uniqueness-of-density 
proof for the time-dependent interacting tunneling problem.

\section{Partition scheme, Hamiltonian, and collision theory}

It is convenient to utilize the partition framework of Caroli {\it et al,}
Ref.~\onlinecite{Caroli}, but retain the full level of details of 
atomistic, many-body calculations everywhere close to the tunneling 
region $\mathbf{r}\sim 0$ at all physically relevant times 
$t\sim 0$ in the collision problem.  For simplicity, the
tunneling structure only comprises a left (right) lead $\mathbf{r}
< z_L (> z_{R}$) plus a central tunneling region `C' in 
$z_L < \mathbf{r} < z_R$.  Atomic units will be used throughout 
and the full kinetic energy is written
\begin{equation}
\sum_s \int d\mathbf{r}' \, \int d\mathbf{r} \,
\hat{\psi}_{s}^{+}(\mathbf{r})
\left(-\frac{1}{2}\{\nabla\delta(\mathbf{r}-\mathbf{r}')\}^2\right) 
\hat{\psi}_{s}(\mathbf{r})
\equiv K_L + K_R + K_C + \delta K
\label{Kdef}
\end{equation}
where $K_{L,R,C}$ has a basis strictly confined to separate 
components and where 
\begin{equation}
\delta K = \sum_{i=L/R}
\sum_s \int d\mathbf{r} \, \hat{\psi}_{s}^{+}(\mathbf{r})
\left(-\frac{1}{2}\{\nabla\, \delta(\mathbf{r}-z_i)\}^2\right)
\hat{\psi}_{s}(\mathbf{r}).
\label{eq:deltaK}
\end{equation}

At time $t \to -\infty$ the partition scheme assumes that each of 
the disconnected subsections of the partitioned system 
$H_d=\sum_{i=L,C,R} H_i$ is in equilibrium at (generally) different 
chemical potentials $\mu_{L/C/R}$.  The operators $N_{L/C/R}$ describe
the electron count in each of the three subcomponents.
Initially, at $t\to -\infty$, the system is described
by a static potential $v_0(\mathbf{r})$ and corresponding
operator 
\begin{equation}
V_0=\int d\mathbf{r} \, v_0(\mathbf{r})\, \hat{n}(\mathbf{r})
\label{eq:V0def}
\end{equation}
where $\hat{n}(\mathbf{r})\equiv\sum_s\hat{\psi}_{s}^{+}(\mathbf{r})
\hat{\psi}_{s}(\mathbf{r})$ denotes the electron-density operator.
I assume, for simplicity, that $v_0(\mathbf{r})$ reduces to a 
uniform background potential $\phi_{L/R}$ (with a value set by the
average electron concentration) far in the leads.
The choice of initial Hamiltonian, 
\begin{equation}
H_d =  
\sum_{i=L/R/C} K_i +V_{0}
\label{eq:Hdterm}\\
\end{equation}
ensures an automatic charge neutrality at $t\to -\infty$
(and at $z\to \pm \infty$ even at finite $t$).  The
equilibrium distribution at $t\to -\infty$ is specified
by a Gibbs energy weighting $H_d-\mu_L N_L -\mu_R N_R - 
\mu_C N_C$ which is independent of the value of the applied 
bias $\phi_{\rm bias} \equiv \mu_L-\mu_R$ and which
exclusively depends on the initial electron concentration in 
the leads (and on the initial electron occupation of the
central island `C').

The LSC-DFT further assumes an adiabatic turning on of 
the tunneling, the electron-electron interaction $W$, 
and of the static electron-scattering potential 
$v_{\rm sc}(\mathbf{r})$ which includes the effects of 
the applied bias and of the set of atomic potentials. One
can also allow for an addition of a time-dependent potential 
$\phi_{\rm g}(\mathbf{r},t>t_0)$ that describes a possible 
gate operation starting at some finite time $t_0$.  The 
time-dependent collision potential is
\begin{equation}
v_{\rm col}(\mathbf{r},t)= [v_{\rm sc}(\mathbf{r})
+ \phi_{\rm g}(\mathbf{r},t)]\exp(\eta t)
\label{eq:vcoldef}
\end{equation}
and the collision problem is formally described by the Hamiltonian
\begin{eqnarray}
H(t) & = & H_d + H_1(t)
\label{eq:Hamilton}\\
H_1(t) & = & 
[ \delta K + W ] \exp(\eta t)+\delta V(t) 
\label{eq:H1term}\\
V_{\rm col}(t)= \delta V(t) & = & 
\int d\mathbf{r} \, v_{\rm col}(\mathbf{r},t) \, \hat{n}(\mathbf{r})
\label{eq:Vterm}
\end{eqnarray}
where the factor $\exp(\eta t), 
\eta\to0^{+}$ expresses the adiabatic turning on.  The collision
term that drives the dynamics is $H_1(t)$.  One may, 
without lack of generality, assume that the collision
potential $v_{\rm col}(\mathbf{r},t)$ also contains an implicit 
multiplicative factor that restricts the basis to a finite, but 
very large region (much larger than $z_{\rm L} < z < z_{\rm R}$);
given the choice for $v_0(|\mathbf{r}|\to\infty)$ this assumption
simply amounts to treating the remote part of the leads as jellium.

The expectation value of the electron density is defined~\cite{Langreth}
\begin{eqnarray}
n(\mathbf{r},t) &=& \langle \hat{n}(\mathbf{r}) \rangle(t) 
\equiv \frac{\hbox{Tr}\{\hat{\rho_0}\, \hat{n}_H(\mathbf{r};t)\}}
{\hbox{Tr}\{\rho_0\}}
\label{eq:neqqsmdef}\\
\quad \hat{n}_H(\mathbf{r};t) & \equiv & \hat{U}(t,-\infty)^{+} \, \hat{n}(\mathbf{r}) \, \hat{U}(t,-\infty)
\label{eq:AHneqqsmdef}
\end{eqnarray}
by establishing the initial (equilibrium) density matrix 
$\hat{\rho}(t\to -\infty) = \hat{\rho}_0$ and by formally solving for 
the ensuing (nonequilibrium) dynamics described by~\cite{LippSchwing} 
\begin{equation}
i\frac{\partial \hat{U}(t,-\infty)}{\partial t} = H(t) \hat{U}(t,-\infty).
\label{eq:genQM}
\end{equation}
This applies to general quantum-statistical problems but it is important
to provide consistent (conserving) approximations to the time-evolution 
$\hat{U}(t,-\infty)$ of the systems. Such approximations are notoriously 
difficult to obtain for interacting systems out of equilibrium.

In this paper, I use the LS many-body collision
theory~\cite{LippSchwing,Pirenne,Gellmann} and describe the 
interacting tunneling problem as a complex collision 
problem~\cite{AlbertXIX29,Yang} in which (an ensemble of) 
fully-interacting many-body electronic states of the 
leads encounter and scatter off a potential in some central tunneling 
region.  For any initial many-body state $|\Phi_{\xi}\rangle$ (eigenstate 
of $H_d$) one may formally obtain the many-body collision state~\cite{LippSchwing}
\begin{equation}
|\Psi_{\xi,+}(t)\rangle = \hat{U}(t,-\infty) | \Phi_\xi\rangle
\label{eq:scatdef}
\end{equation}
from a direct application of the temporal-evolution operator~(\ref{eq:genQM}).
The case of a purely static scattering potential is then described as 
an elastic many-body collision but the finite bias still causes actual 
electron-electron scattering~\cite{QCLdisc} that invalidates
assumptions of ballistic transport.  An inelastic collision event 
arises when the ensemble of many-body states scatters off a 
corresponding time-dependent collision potential.\cite{acKondo,philal}  
An effective time-dependent potential may also arise in the 
description of tunneling in the presence of a quantized boson 
field,\cite{Phon,FloRun,PHexcit,RecentBrandbyge}
as is relevant for further investigations of optoelectronic devices,
for example, lasers under typical operating conditions.

The many-body collision problem (for tunneling) is from the outset 
specified by the choice of partitioning, that is, (i) a specific 
choice of Hamiltonian $H_d$ with an interrupted kinetic energy 
$K-\delta K$ and (ii) the lead electron concentration, and (iii) 
the choice of an initial state $|\Phi_0\rangle$ (or ensemble 
of initial states, $\hat{\rho}_0$) that also formally depends 
on the initial distribution in the central region `C'. 

\section{Density functional theory of interacting tunneling}

For a time-dependent collision potential it is relatively simple to 
generalize the TD-DFT analysis\cite{RungeGross,InftyTDDFT} to the 
many-body collision theory of nonequilibrium interacting
tunneling, appendix A.  This demonstrates that the time-dependent 
density is a unique functional of the time-dependent collision potential.
The time-dependent collision density functional theory contains the 
steady-state formulation of LSC-DFT as a limiting case under some
conditions.

\subsection{Density functional theory of time-dependent tunneling}

Applying a time-dependent gate voltage to a nonequilibrium tunneling
system probes a response~\cite{acKondo,philal} that
reflects electron correlations. It is interesting in itself to 
develop a formal density functional theory basis for such 
time-dependent many-body collision problems.

For a given bias and partitioning of the general time-dependent 
collision problem I analyze the mapping  $\mathcal{N}: 
v_{\rm col}(\mathbf{r},t) \longrightarrow n(\mathbf{r},t)$ and 
argue
\begin{itemize}

\item[{\it O1.}] {\it The electron density $n(\mathbf{r},t)$
is a unique functional of the collision potential
$v_{\rm col}(\mathbf{r},t)$.}

\end{itemize}
For the collision problem the goal is to determine the nonequilibrium 
time-evolution of a single state (or a grand-canonical ensemble) 
of an infinite open system under a collision term $H_1(t)$ 
which includes an adiabatic turning on of both tunneling $\delta K$, 
the electron-electron interaction $W$, and a time-dependent potential 
$\delta V(t)$ (with a finite basis).  In contrast, TD-DFT considers a 
single state which from the outset is described by $K+W$ and which 
evolves under an external potential $V(t)$ and demonstrates
uniqueness  of the time-dependent density when the system is 
finite~\cite{RungeGross} and/or when the external potential $V(t)$ 
has a finite basis.\cite{InftyTDDFT} However, it is straightforward 
to generalize the reductio ad absurdum argument from TD-DFT to
the present many-body collision formulation of interacting 
nonequilibrium tunneling described in a partition scheme. This is 
because the {\it difference\/} of collision terms, $H_1(t)-H_1'(t)$, 
remains exclusively specified by the difference in collision potentials, 
$v_{\rm col} (\mathbf{r},t)- v_{\rm col}'(\mathbf{r},t)$ (and this
difference may be assumed to have a finite basis).

Formally the observation {\it O1\/} only establishes the uniqueness of 
the density variation and history for a given bias and for a given
choice of partition scheme (that is, choice of $z_{\rm L/R}$ and 
choice of the initial occupation in the central region `C'). 
A different partition scheme produces a different unique mapping,
$\tilde{\mathcal{N}}$ (as well as corresponding T-matrix functionals
for the scattering behavior).  However, the choice of partition scheme
must become irrelevant for very large $z_{\rm R}-z_{\rm L}$ and 
basis of the collision term $\delta V(t)$. This is argued on 
physical grounds for tunneling systems that lack singular responses:
since dephasing eventually decouples the time and spatial 
correlations~\cite{NEQreform} the solution $n(\mathbf{r},t)$ 
must eventually become insensitive to memory 
effects as well as details of the potential far in the 
leads.\cite{Stefanucci,TranSiesta}

\subsection{Density functional theory of steady-state tunneling}

For static collision problems the four-dimensional variational 
space of a general density history $n(\mathbf{r},t)$ naturally
becomes overcomplete in its definition of the scattering
potential $v_{\rm sc}(\mathbf{r})$.  The mapping 
$\mathcal{N}^{-1}: n(\mathbf{r},t) \longrightarrow v_{\rm col} 
(\mathbf{r},t)$ shows that a general density history causes 
potential variations
\begin{equation}
\delta v  =  \frac{\delta \mathcal{N}^{-1}}{\delta n} \delta n(\mathbf{r},t)
=  \frac{\delta \mathcal{N}^{-1}}{\delta n} 
\left[\nabla n \delta \mathbf{r}
+ \frac{\partial n}{\partial t} \delta t\right]
\label{eq:densrestrict}
\end{equation}
that are incompatible with the steady-state transport assumption.  

For time-independent collision problems that have a steady-state 
solution, $n(\mathbf{r},t)= n(\mathbf{r},0)$, I argue
\begin{itemize}

\item[{\it O2.}] {\it The time-independent scattering potential
$v_{\rm sc}(\mathbf{r})$ is uniquely determined by the
steady-state density $n(\mathbf{r})\equiv n(\mathbf{r},0)$.}

\end{itemize}
The adiabatic turning on of the static scattering potential 
$v_{\rm sc}(\mathbf{r})$ can be viewed as a limiting case of 
a time-dependent tunneling problem with the restricted variation:
\begin{equation}
\delta v_{\rm col} = \nabla v_{\rm sc}\, \exp(\eta t)\, 
\delta \mathbf{r} + \eta \, v_{\rm sc} \, \exp(\eta t)\, \delta t 
\longrightarrow 
\nabla v_{\rm sc}\, \delta \mathbf{r}.
\label{eq:potlimit}
\end{equation}
The unique mapping $\mathcal{N}: v_{\rm col} (\mathbf{r},t) 
\longrightarrow n(\mathbf{r},t) $ identifies the corresponding set 
of relevant density variations. If the steady-state
tunneling problem is characterized by non-divergent values
of $\delta\mathcal{N}/\delta v$ we have\cite{AvalancheDiode}
\begin{equation}
\frac{\partial n}{\partial t} = 
\frac{\delta\mathcal{N}}{\delta v}\, \eta\, v_{\rm sc} \, \exp(\eta t) 
\longrightarrow 0.
\label{eq:relevant}
\end{equation}
This is the condition that specifies the steady-state scattering 
solutions, $\delta n(\mathbf{r})= \delta n(\mathbf{r},t=0)$.

The observation {\it O2\/} permits formulation of universal density
functionals that characterize variational expressions for the 
noninteracting and interacting T-matrices in LSC-DFT.
Formally these universal functionals also depend on the 
assumptions that are build into the partition scheme.  For some 
tunneling problems it may be important to retain a functional 
dependence on the initial island occupation $\mu_C$.  Here I assume 
that the solution of the steady-state LSC-DFT problem is a functional 
only of the density (for a given applied bias and choice of $\mu_L$ 
and $\mu_R$).


\section{A Lippmann-Schwinger T-matrix functional description}

Formulation of a single-particle scheme with independent
dynamics of fictitious particles is important for an accurate
and efficient evaluation of the electron density in interacting 
nonequilibrium tunneling problems.  Prerequisites for such a formulation 
are (a) corresponding variational expressions of a many-body quantity 
evaluated both in the interacting and noninteracting cases, and
(b) universal density functionals to characterize those variational
physical quantities. The LSC-DFT uses the many-body LS variation
principles for the interacting and noninteracting many-body T-matrices 
as well as exact reformulations to satisfy those necessary conditions.
The LS variational properties also permit the LSC-DFT to specify an 
exact implicit equation for the nonequilibrium electron density.

\subsection{The Lippmann-Schwinger variation principle}

In their seminal paper~\cite{LippSchwing} Lippmann and Schwinger
identified a set of variational properties for the collision 
problems. The original theory ignores the self-energy shifts 
associated with the adiabatic tuning on of the collision term 
$H_1(t)$ but the work was soon after supplemented and regularized
by additional analysis.\cite{Pirenne,Gellmann}  The LS variational 
principle~\cite{Albert} applies for any combination of collision 
states and it should be straightforward to generalize the following 
also to finite-temperature tunneling problems.  Like 
in traditional equilibrium DFT, however, the focus will likely 
remain on zero-temperature properties.  It is natural 
to build a functional that reflects the evolution of the 
ground-state $|\Phi_0\rangle$ of the original 
disconnected system. 

The LSC-DFT provides an exact variational evaluation of the 
ground-state-to-ground-state transition matrix element $T_V[n]$, 
with usual definition $T_V[n]=\langle \Phi_0| H_1| \Psi_{0,+}[n]\rangle$. 
This matrix element is a functional of the tunneling electron density 
because the scattering state $|\Psi_{0,+}[n]\rangle$ is specified by 
the choice of external potential. Through the optical 
theorem,\cite{AlbertOptTheorem} this T-matrix element 
characterizes the total rate of tunneling (a charge transfer process)
arising in the presence of full electron-electron interaction.  

The LS variational principle is expressed using the notation of
Ref.~\onlinecite{LippSchwing}. I generally follow the discussion 
in Ref.~\onlinecite{Albert} and introduce $G_{\pm}^{d}=(E_0-H_d 
\pm i\eta)^{-1}$ as the retarded ($+$) and advanced ($-$) Green 
function operator 
of the original disconnected Hamiltonian $H_d$ while 
\begin{equation}
|\Psi_{0,\pm}[n]\rangle = |\Phi_0\rangle + 
G_{\pm}^d H_1 |\Psi_{0,\pm}[n]\rangle
\label{eq:tradLSequation}
\end{equation}
identifies the many-body (collision) state 
that evolves forward/backwards in time under the collision
term $H_1$.  These states are functionals 
exclusively of the density $n(\mathbf{r})$ (or 
$n(\mathbf{r},t)$ in the wider problems beyond the
present scope). I further introduce 
\begin{eqnarray}
\Xi_{-}[n,V] & \equiv & \langle \Psi_{0,-}[n]|H_1|
\Phi_0\rangle,
\label{eq:XimFormT}\\
\Xi_{+}[n,V] & \equiv & \langle \Phi_{0}|H_1|
\Psi_{0,+}[n]\rangle,
\label{eq:XipFormT}\\
\Upsilon[n,V] & \equiv & 
\langle \Psi_{0,-}[n]|H_1-H_1 G_+^d H_1| \Psi_{0,+}[n]\rangle
\label{eq:UpsFormT}
\end{eqnarray}
which, like the compensated form
\begin{equation}
T_V[n]= \Xi_{-} [n,V] + \Xi_{+} [n,V]
- \Upsilon[n,V],
\label{eq:Tvarform}
\end{equation}
are functionals of the density $n$ but also contain an explicit
dependence on $V=V_0+\delta V$ through the collision term
$H_1$. All four functionals represent a correct evaluation of 
the T-matrix behavior when evaluated at the correct density 
$n$ (the density $n$ that results under the collision term 
$H_1$). For $\Upsilon[n,V]$ this follows by a simple application 
of the Dyson equation, $T_+=1-H_1 G_+^d=1-G_+^dH_1$, see 
Refs.~\onlinecite{LippSchwing,AlbertXIX29}. 

The key observation for the LSC-DFT formulation is that the
extremum identified by the variational condition
\begin{equation}
\frac{\delta T_V[n]}{\delta n} = 
\frac{\delta T_V}{\delta \Psi_{0,-}}\frac{\delta \Psi_{0,-}}{\delta n}
+\frac{\delta T_V}{\delta \Psi_{0,+}}\frac{\delta \Psi_{0,+}}{\delta n}=0
\label{eq:Tmatderiv}
\end{equation}
identifies the electron density $n$ that solves the collision 
problem $H_1$.  This follows from the (many-body) LS variational 
principle~\cite{LippSchwing,AlbertXIX29} because 
the derivative $\delta T_V/\delta \Psi_{0,-(+)}$
is directly proportional to the many-body LS equation for scattering
states $|\Psi_{0,+(-)}\rangle$, Ref.~\onlinecite{LippSchwing}. 

The noninteraction collision problem, defined by $H_{1}^{0}
=\delta K+\delta V$, has a corresponding density functional
description. It has a different unique mapping between
the density and the potential $\delta V$ and different
scattering states $|\Psi^0_{0,\pm}\rangle$ and this mapping 
defines other (related) functionals 
\begin{eqnarray}
\Xi^0_{-}[n,V] & \equiv & \langle \Psi^0_{0,-}[n]|H_1^0|
\Phi_0\rangle,
\label{eq:XimFormT0}\\
\Xi^0_{+}[n,V] & \equiv & \langle \Phi_{0}|H_1^0|
\Psi^0_{0,+}[n]\rangle,
\label{eq:XipFormT0}\\
\Upsilon^0[n,V] & \equiv & 
\langle \Psi^0_{0,-}[n]|H_1^0-H_1^0 G_+^d H_1^0| \Psi^0_{0,+}[n]\rangle
\label{eq:UpsFormT0}\\
T_V^0[n] &=& \Xi^0_{-} [n,V] + \Xi^0_{+} [n,V]
- \Upsilon^0[n,V].
\label{eq:T0varform}
\end{eqnarray}
The extremum, identified by the variational condition 
\begin{equation}
\delta T_V^0[n]=0,
\label{eq:T0matderiv}
\end{equation}
identifies the density that solves the noninteracting problem
$H_1^0=\delta K+ \delta V$. 

\subsection{Universality of T-matrix functionals in LSC-DFT}

To obtain a description given in terms of universal functionals
it is necessary to identify the partial contributions that arise 
from the kinetic-energy addition, the electron-electron interaction, 
and the potential scattering and to find a method to evaluate the 
difficult many-electron effects once and for all.  This is possible 
by formal manipulation using the Dyson equation and 
by use of the LS equation itself.  By 
construction, the formal manipulation does not invalidate the 
underlying variational character of the 
LS formulation (\ref{eq:Tvarform}).

I first introduce scattering states and Green function operators 
for the set of partial collision problems defined at $\delta V\equiv 0$.
I use $|\chi_{0,\pm}^0\rangle$ and $|\chi_{0,\pm}\rangle$ 
to denote the collision states for the $\delta V\equiv 0$
noninteracting and interacting connected problems, 
given by $H_d+\delta K$ and $H_d+\delta K + W$, respectively.
Also, $G_{\pm}^0=[E_0-(H_d+\delta K)\pm i\eta]^{-1}$ 
and $G_{\pm}=[E_0-(H_d+\delta K+W)\pm i\eta]^{-1}$ identify 
the noninteracting and interacting Green function operator 
at $\delta V=0$.  These noninteracting and interacting collision
problems contain an implicit choice of potential $V_0$ while
the general noninteracting and interacting collision problem
is described by $V=V_0+\delta V\equiv V_{\rm col}$. Neither
$\chi_{0,\pm}$ nor $\chi^0_{0,\pm}$ are therefore functionals
of $n$ (and the same applies for all Green functions in use).

For the matrix element $\Xi_{+}^0[n]=\langle
\Phi_0|(\delta K + \delta V) |\Psi_{0,+}^0[n]\rangle$ I use a 
simple formal manipulation of the `ket' state
\begin{eqnarray}
|\Psi_{0,+}^0[n]\rangle & = & |\Phi_0\rangle 
+ G_+^d (\delta K + \delta V) |\Psi_{0,+}^0[n]\rangle 
\nonumber\\
& = & |\chi_{0,+}^0\rangle + G_+^0 \delta V |\Psi_{0,+}^0[n]\rangle,
\label{eq:LSreform}
\end{eqnarray}
as can be verified by a direct application of the 
LS equation.\cite{Albert} The resulting separation
\begin{equation}
\Xi_{+}^0[n,V] = \langle \Phi_0 | \delta K |\chi_{0,+}^0\rangle
+ \langle \chi_{0,-}^0 | \delta V | \Psi_{0,+}^0[n]\rangle
\label{eq:A0reform}
\end{equation}
can, of course, be repeated for a separation also of 
$\Xi_{-}^0[n]$ and of the corresponding interacting expression 
$\Xi_{\pm}[n,V]$, for example,
\begin{equation}
\Xi_{+}[n,V] = \langle \Phi_0 | (\delta K+W) |\chi_{0,+}\rangle
+ \langle \chi_{0,-} | \delta V | \Psi_{0,+}[n]\rangle.
\label{eq:Areform}
\end{equation}
This reformulation can also be derived, Eqs. (XIX.9) and 
(XIX.120) of Ref.~\onlinecite{Albert}, by applying the Green theorem 
on the weighted overlap between the two collision states. The Green
theorem plays a central role in NEGF calculations for the open 
tunneling systems and enters, for example, in the quantum-kinetic
based derivation of resonant-tunneling rate equations.\cite{RateEq}

For the noninteracting matrix element $\Upsilon^0[n,V]=
\langle \Psi_{0,-}^0[n]| (\delta K + \delta V)
\{1-G_+^d(\delta K +\delta V)\} 
|\Psi_{0,+}^0[n]\rangle $ (and for $\Upsilon[n,V]$) it is 
necessary to first expand the `bra' and `ket' 
collision states by the LS equation.  I collect 
terms involving either a $\langle \chi_{0,-}^0|$ or
a $\langle \Psi_{0,-}^0[n]|$ `bra' state and either a
$| \chi_{0,+}^0 \rangle$ or a $| \Psi_{0,+}^0[n] \rangle$
`ket' state separately and I use the underlying 
quantum-kinetic (Dyson) equation for simplification.  
Taking one of the cross terms as an example, one obtains
\begin{equation}
\langle \chi_{0,-}^0| \delta K 
\{G_+^0[V_0]-G_+^d \delta K G_+^0[V_0] - G_+^d \} \delta V 
| \Psi_{0,+}^0[n] \rangle \equiv 0.
\label{eq:CTvanish}
\end{equation}
Repeated applications of the Dyson equation completes
the separation
\begin{equation}
\Upsilon^0[n,V]  =  
\langle \chi_{0,-}^0 |
\delta K \{1 - G_+^d \delta K \} | \chi_{0,+}^0 \rangle
+ \langle \Psi_{0,-}^0[n] |
\delta V \{1 - G_+^0 \delta V \} | \Psi_{0,+}^0[n] \rangle.
\label{eq:Ups0reform}
\end{equation}
A corresponding expansion applies, of course, also for the 
interacting matrix element
\begin{equation}
\Upsilon[n,V]  =  
\langle \chi_{0,-} |
(\delta K+W) \{1 - G_+^d (\delta K+W) \} | \chi_{0,+} \rangle
+ \langle \Psi_{0,-}[n] |
\delta V \{1 - G_+ \delta V \} | \Psi_{0,+}[n] \rangle.
\label{eq:Upsreform}
\end{equation}

The electron-electron interaction effects on the many-body scattering 
problem can now be expressed in universal functionals.
A set of complex constants 
\begin{eqnarray}
{a}_{-} & \equiv & \langle \chi_{0,-}| (\delta K+W) |\Phi_0 \rangle
\label{eq:conAmdef}\\
{a}_{+} & \equiv & \langle \Phi_0 | (\delta K + W) | \chi_{0,+} \rangle
\label{eq:conApdef}\\
{b} & \equiv & \langle \chi_{0,-}| (\delta K +W)[1-G_+^d (\delta K +W)]
                                  | \chi_{0,+} \rangle
\label{eq:conbdef}
\end{eqnarray}
(along with corresponding definitions $a_{\pm}^0, b^0$ for the 
noninteracting case) characterizes the dynamics in the
absence of the collision potential (at $\delta V=0)$.  More importantly,
a set of collision-state matrix elements
\begin{eqnarray}
\mathcal{A}_{-}[n](\mathbf{r}) & \equiv & 
            \langle \Psi_{0,-}[n]| \hat{n}(\mathbf{r}) 
|\chi_{0,+} \rangle
\label{eq:mathAmdef}\\
\mathcal{A}_{+}[n](\mathbf{r}) & \equiv & 
          \langle \chi_{0,-} | \hat{n}(\mathbf{r}) 
| \Psi_{0,+}[n] \rangle
\label{eq:mathApdef}\\
\mathcal{B}_1[n](\mathbf{r}) & \equiv & 
          \langle \Psi_{0,-}[n]| \hat{n}(\mathbf{r})
                                  | \Psi_{0,+}[n] \rangle
\label{eq:mathBdef}\\
\mathcal{B}_2[n](\mathbf{r},\mathbf{r}') & \equiv & 
          \langle \Psi_{0,-}[n]| 
                    \hat{n}(\mathbf{r}) G_+
                    \hat{n}(\mathbf{r}')
                                  | \Psi_{0,+}[n] \rangle
\label{eq:mathB2def}
\end{eqnarray}
represents universal density functionals that determine the many-body
dynamics when the collision potential is included in the presence of 
full electron-electron interaction (while corresponding universal 
functionals $\mathcal{A}_{\pm}^0[n]$ and $\mathcal{B}_{1,2}^0[n]$
characterize the full collision problem at $W=0$). 
The variational form of the interacting and noninteracting 
T-matrices can thus be reformulated
\begin{eqnarray}
T_V[n] & = & a_{-}+a_{+}-b 
+ \int \, d\mathbf{r} \, v_{\rm sc}(\mathbf{r}) \,
\mathcal{K}_V[n](\mathbf{r})
\label{eq:Tvarreform}\\
T_V^{0}[n] & = & a_{-}^{0}+a_{+}^{0}-b^{0} 
+ \int \, d\mathbf{r} \,  v_{\rm sc}(\mathbf{r}) \,
\mathcal{K}_V^{0}[n](\mathbf{r})
\label{eq:Tvar0reform}\\
\mathcal{K}_V^{(0)}[n](\mathbf{r}) & = & 
\mathcal{A}_+^{(0)}[n](\mathbf{r}) 
+ \mathcal{A}_-^{(0)}[n](\mathbf{r}) 
- \mathcal{B}_1^{(0)}[n](\mathbf{r}) 
+ \int d\mathbf{r}' \,
\mathcal{B}_2^{(0)}[n](\mathbf{r},\mathbf{r}') 
 v_{\rm sc}(\mathbf{r}').
\label{eq:mathcalKdef}
\end{eqnarray}
In essence, calculation of a set of universal functionals 
(for relevant choices of chemical potentials $\mu_{\rm L/R}$) permits a
simple general evaluation of the interacting T-matrix for arbitrary 
scattering potentials $v_{\rm sc}(\mathbf{r})$.  

The LSC-DFT description also permits a succinct formulation of 
the interaction effects on the T-matrix functional derivatives
\begin{equation}
\frac{\delta (T_V[n]-T_V^0[n])}{\delta n(\mathbf{r})}= 
 v_{\rm sc}(\mathbf{r}) 
\frac{\delta (\mathcal{K}_V[n]-\mathcal{K}_V^0[n])}
{\delta n(\mathbf{r})}\equiv 
 v_{\rm sc}(\mathbf{r}) \frac{\delta \Delta \mathcal{K}_V[n]}
{\delta n(\mathbf{r})}.
\label{eq:funcderiv}
\end{equation}
The interaction effect is expressed as a complex function of 
$\mathbf{r}$ and it is entirely specified by suitable
approximations to universal density functionals.  

\subsection{Variational solution of the interacting collision problem}

Separating out the noninteracting dynamics (for which we can seek
highly accurate characterizations) and the interaction 
effect (\ref{eq:funcderiv}) defines a formal LSC-DFT solution
\begin{equation}
\frac{\delta T_V[n]}{\delta n(\mathbf{r})}=
\frac{\delta T^0_V[n]}{\delta n(\mathbf{r})}
+ v_{\rm sc}(\mathbf{r}) \frac{\delta 
\Delta \mathcal{K}_V[n]}{\delta n(\mathbf{r})},
\label{eq:formalSolve}
\end{equation}
which constitutes an exact but implicit equation for 
the nonequilibrium electron density.

The formal LSC-DFT solution~(\ref{eq:formalSolve}) serves as 
a natural starting point for search for a single-particle 
scheme for calculations of the density in specific nonequilibrium 
tunneling systems.  The possibility is exciting, for a rigorous 
single-particle scheme would permit efficient and exact calculations 
of $\delta T_V^0[n]$ and would ensure automatic current conservation 
in {\it ab initio\/} calculations specified by universal functionals.  
The single-particle LS equation certainly determines the noninteracting 
many-body dynamics described by $\delta T_V^0[n]/\delta n$.  
It is not a priori clear that the interaction term 
(\ref{eq:funcderiv}) represents an additional effect caused by 
some {\it effective single-particle scattering\/} and it is not a 
priori clear that a single-particle scheme exists for 
the LSC-DFT. However, the present results show that the LSC-DFT
satisfies necessary conditions and motivate a search for rigorous
single-particle formulations.   

\section{Summary and outlook}

This paper formulates a Lippmann-Schwinger collision density functional
theory (LSC-DFT) for nonequilibrium interacting steady-state tunneling.  
The theory rests on the Lippmann-Schwinger variational principle for 
the interacting and noninteracting many-body T-matrices and includes
exact reformulations that express the variational T-matrix forms 
through universal density functionals. The variational property of
the LSC-DFT specifies an exact implicit equation~(\ref{eq:formalSolve}) 
for the electron density.  The LSC-DFT furthermore fulfills {\it 
necessary\/} conditions for a possible formulation of a rigorous 
single-particle scheme. The present results motivate a future study 
(using the formal LSC-DFT solution (\ref{eq:formalSolve}) and the LS 
variational properties of single-particle scattering) to explore 
conditions on the dynamics and to test if a rigorous single-particle 
scheme can be defined in LSC-DFT.

Of course, any implementation of a LSC-DFT method must also rely on
successful formulation of a good approximation for the universal 
functionals $\mathcal{A}_{\pm}[n]$ and $\mathcal{B}_{1,2}[n]$ 
that characterize the complex many-electron collision behavior.
The formulation of TD-DFT-based {\it ab initio\/} 
calculations\cite{Burke,Stefanucci} facilitates a program 
to explore the T-matrix behavior for a range of scattering 
potentials and thereby deconvolute approximations for the 
universal functionals $\mathcal{A}_{\pm}[n]$ and $\mathcal{B}_{1,2}[n]$.
The partition-scheme method for time propagation of the tunneling 
many-body wavefunction~\cite{TDDFTpartitionCalcs} may allow a 
direct extraction of T-matrices and simplify the task.
It is natural to first seek functionals that have a local-density flavor
in the parameterization of Eqs.~(\ref{eq:mathAmdef})-(\ref{eq:mathB2def})
but it is also possible that more complex functional forms must
be explored.  Exact solutions of nonequilibrium correlated-electron 
model systems~\cite{exactKondo} present possibilities for further
refining parameterizations of approximations for the universal 
functionals.  

\section{acknowledgments}

The author acknowledges discussions with B.~I.~Lundqvist. Research 
supported by the Swedish Research Council (VR) and by the Swedish 
Government agency for Innovation Systems (VINNOVA). 


\appendix

\section{Uniqueness of density in collision theory}

I argue uniqueness of the time-dependent density {\it O1\/} 
for the complex LS collision problem of open tunneling systems
in the partition scheme as a relatively straightforward generalization of 
the TD-DFT analysis for finite systems~\cite{RungeGross} and for
infinite systems with a restricted basis of the single-particle
potential.\cite{InftyTDDFT}  
The grand-canonical foundation, the use of the partition scheme, 
and the basis in quantum-kinetic equations makes the analysis of
the many-body collision problem slightly different from that of 
Ref.~\onlinecite{InftyTDDFT} and the argument is included here 
for completeness.  

For a specific partition with given initial configuration and 
initial density matrix operator $\hat{\rho}(t\to -\infty)=
\hat{\rho}_0$, I consider the time evolution 
\begin{equation}
i\frac{\partial\hat{\rho}(t)}{\partial t}=\left[H(t),\hat{\rho}(t)\right],
\label{eq:rhotevo}
\end{equation}
with formal solution given by the many-body evolution operator 
\begin{equation}
\hat{\rho}(t)=\hat{U}(t,-\infty)^{+}\hat{\rho}_0 \hat{U}(t,-\infty).
\end{equation}
I compare two similar systems given by $H_1(t)$ and $H_1'(t)$ 
for which the $k$'th time-derivative of the collision potential 
begins to differ at some time $t_i$, Eq.~(3) of 
Ref.~\onlinecite{RungeGross}. The current-density operator 
\begin{equation}
\hat{j}(\mathbf{r})=
(2i)^{-1}\sum_s [\nabla \hat{\psi}_s^{+}(\mathbf{r})]\psi_s(\mathbf{r}) -
\hat{\psi}_s^{+}(\mathbf{r})[\nabla \psi_s(\mathbf{r})]
\label{eq:jopdef}
\end{equation}
constitutes a sensitive probe of system differences at
times immediately thereafter, $t=t_i^{+}$. 

The density matrices $\hat{\rho}(t)$ and $\hat{\rho}'(t)$ that 
correspond to $H_1(t)$ and $H_1(t)$ must, of course, agree at $t_i$.
Use of (\ref{eq:rhotevo}) permits the evaluation
\begin{eqnarray}
i\frac{\partial}{\partial t} \Delta j(\mathbf{r},t_i) & = &
\hbox{Tr} \left\{\left[H_1(t_i)-H_1'(t_i), \hat{\rho}(t_i)\right]
\hat{j}(\mathbf{r})\right\} 
\nonumber\\
& = & \int \, d\mathbf{r}' \, \left(v_{\rm col}(\mathbf{r}',t_i)-
v_{\rm col}'(\mathbf{r}',t_i)\right) \, 
\hbox{Tr} \left\{\hat{\rho}(t_i)
\left[\hat{j}(\mathbf{r}),\hat{n}(\mathbf{r'})\right]
\right\}
\nonumber\\
& = & i \hbox{Tr} \left\{\hat{\rho}(t_i)\hat{n}(\mathbf{r})\right\}
\nabla
[v_{\rm col}(\mathbf{r},t_i)-v_{\rm col}'(\mathbf{r},t_i)]
\label{eq:jdiff}
\end{eqnarray}
of the system differences in time-evolution of the current 
expectation values.  The third line of (\ref{eq:jdiff})
results from an operator identity and by partial integration
in formal manipulations that directly mirror those of the TD-DFT analysis.
It applies because the finite (but assumed very large) basis for 
$v_{\rm col}(\mathbf{r},t)$ and for $v_{\rm col}'(\mathbf{r},t)$ 
eliminates surface contributions.  

If the potentials themselves differ at $t_i$ it follows directly 
that the current densities must differ at a time immediately 
thereafter. If instead the potentials only
differ at some derivative of order $k\geq 1$ we proceed by
direct differentiation of Eq.~(\ref{eq:jdiff}):
\begin{equation}
i\frac{\partial}{\partial t}
\left(i\frac{\partial}{\partial t}\right)^{k} 
\Delta j(\mathbf{r},t_i) = i \hbox{Tr} 
\left\{\hat{\rho}(t_i)\hat{n}(\mathbf{r})\right\}
\nabla \left(i\frac{\partial}{\partial t}\right)^{k}
[v_{\rm col}(\mathbf{r},t_i)-v_{\rm col}'(\mathbf{r},t_i)]\neq 0.
\label{eq:kthderivresult}
\end{equation}
It follows that the current densities must
differ at time $t_i^{+}$.

Finally, uniqueness of the electron density $n(\mathbf{r},t)$ 
results by direct application of the reductio ad absurdum argument 
given for infinite-system TD-DFT in Ref.~\onlinecite{InftyTDDFT}.  
Using $n(\mathbf{r},t_i)= 
\hbox{Tr}\{\hat{\rho}(t)\hat{n}(\mathbf{r})\}$ and the 
continuity equation gives
\begin{equation}
\frac{\partial^{k+2}}{\partial t^{k+2}} 
\left[ n(\mathbf{r},t_i)-n'(\mathbf{r},t_i)\right]
  =  -\nabla\cdot \left[n(\mathbf{r},t) \nabla u(\mathbf{r},t_i)\right],
\label{eq:difconteq}
\end{equation}
\begin{equation}
u(\mathbf{r},t_i)  =  \frac{\partial^k}{\partial t^k}
[v_{\rm col}(\mathbf{r},t_i)-v_{\rm col}'(\mathbf{r},t_i)],
\label{eq:udef}
\end{equation}
that is, in agreement with Eq.~(6) of Ref.~\onlinecite{RungeGross}.
Use of Green's first identity shows that
\begin{equation}
- \int d\mathbf{r} \, u(\mathbf{r},t_i) 
\, \nabla \cdot \left[n(\mathbf{r},t) \nabla u\right]
=
\int d\mathbf{r} \, 
n(\mathbf{r},t_i)[\nabla u(\mathbf{r},t_i)]^2 
\label{eq:GreenFI}
\end{equation}
because our collision problem permits us to make an implicit 
assumption of a finite basis for $v_{\rm col}(\mathbf{r},t)$ so
that $u(|\mathbf{r}|\to\infty)=0$.
As in Refs.~\onlinecite{RungeGross,InftyTDDFT} it follows
that the difference (\ref{eq:difconteq}) must be nonzero 
since $n(\mathbf{r},t_i) [\nabla u(\mathbf{r},t_i)]^2\geq 0$
and since $n(\mathbf{r},t_i) [\nabla u(\mathbf{r},t_i)]^2$ cannot 
vanish identically.
In summary, a reductio ad absurdum argument shows ({\it O1\/})
that the time-dependent density variation $n(\mathbf{r},t)$ is
a unique functional of the collision potential $v_{\rm col}
(\mathbf{r},t)$ in the many-body LS collision problem that
is used here to describe nonequilibrium interacting tunneling.



\end{document}